\documentclass[preprint,aps,amssymb,showpacs,superscriptaddress,nofootinbib]{revtex4}
\usepackage{graphicx}

\newcommand{\Case}[2]{{\textstyle \frac{#1}{#2}}}
\newcommand{\sgn}{\mathop{\mathrm{sgn}}}

\begin{document}
%
\preprint{IMSc/2005/03/05}

\title{On Energy Conditions and Stability in
 Effective Loop Quantum Cosmology}

\author{Golam Mortuza Hossain}
\email{golam@imsc.res.in}
\affiliation{The Institute of Mathematical Sciences,
CIT Campus, Chennai-600 113, India.}

\begin{abstract}
In isotropic loop quantum cosmology, non-perturbatively modified
dynamics of a {\em minimally} coupled scalar field violates {\em
weak}, {\em strong} and {\em dominant} energy conditions when
they are stated in terms of equation of state parameter. The
violation of strong energy condition helps to have non-singular
evolution by evading singularity theorems thus leading to a
generic inflationary phase. However, the violation of weak and
dominant energy conditions raises concern, as in general
relativity these conditions ensure causality of the system and
stability of vacuum via Hawking-Ellis conservation theorem. It is
shown here that the non-perturbatively modified kinetic term
contributes {\em negative} pressure but {\em positive} energy
density. This crucial feature leads to violation of energy
conditions but ensures {\em positivity} of energy density, as
scalar matter Hamiltonian remains bounded from below. It is also
shown that the modified dynamics restricts {\em group velocity}
for inhomogeneous modes to remain {\em sub-luminal} thus ensuring
causal propagation across spatial distances.
\end{abstract}


\maketitle


\section{Introduction}

In general theory of relativity, dynamics of a spacetime is
influenced by matter stress-energy tensor. Naturally, many
properties regarding spacetime evolution can be concluded
assuming some general properties of the matter stress-energy
tensor, without having to know the details of the individual
contributions from different matter sources. These requirements
on the matter stress-energy tensor, widely called {\em energy
conditions}, have been used to prove several important theorems
in classical general relativity. One such theorem, the
Hawking-Ellis {\em conservation theorem} \cite{Hawking,Wald} says
that if the matter stress-energy tensor is conserved, satisfies
{\em dominant energy condition} and vanishes on a closed, {\em
achronal} set $S$ then it also vanishes in the {\em domain of
dependence} $D(S)$ of the set. Physically, this theorem ensures
the stability of classical vacuum. As mentioned, the conservation
theorem stands true provided the matter stress-energy tensor
satisfies the dominant energy condition.  This condition requires
local energy density to be {\em non-negative} for all time-like
observer and the energy-momentum 4-current to be {\em
non-spacelike} {\em i.e.} the speed of energy-flow should not be
exceeding the speed of light.  Naturally, the violation of
dominant energy condition raises concern about the causality and
the stability of the system.  However, it is worth pointing out
that the above theorem does {\em not} have the {\em converse}
{\em i.e.} although the dominant energy condition satisfying
matter ensures causality and stability of the system but
violation of this condition {\em does not} necessarily imply that
the system violates causality or is unstable (see for example
\cite{McInnes}). In such a situation, these issues should be
considered for the specific context, as dominant energy condition
violation and the Hawking-Ellis conservation theorem no longer
vouch for the causality and the stability of the system.

In the cosmological context, the issue of dominant energy
condition violation has acquired significant importance in recent
literature. The observational evidences
\cite{Riess:SN,Perlmutter:SN} seem to suggest that in our
universe major fraction of the energy density is contributed by
some kind of mysterious {\em dark energy} that exerts {\em
negative} pressure. The experimental data in this context not
only allows but often favours the values of the equation of
state parameter to be less than $-1$ for the dark energy
component
\cite{Caldwell:PM,Melchiorri:DEOS,Alam:DEM,Choudhury:SN,
Wang:DE,Nesseris:SN,Gong:SN,Alam:DDER,Wang:DESN,Feng:DE,
Jassal:DE,Corasaniti:DE}. Such values of the equation of state
parameter require violation of dominant energy condition. This
makes the problem of the dark energy even more severe which is
otherwise itself a major theoretical challenge in the present day
cosmology \cite{Weinberg:CCP,Carroll:CC,Sahni:CPL,Carroll:CCLR,
Peebles:CCDE,Padmanabhan:CCPR}. A popular proposed model for
dominant energy condition violating dark energy is so called {\em
phantom matter}
\cite{Caldwell:PM,Schulz:PDE,Carroll:EOS,Singh:PF,Sami:PF}.  The
phantom matter is essentially a minimally coupled scalar field
model but with relatively {\em negative} kinetic term ( but see
\cite{Onemli:SA,Nojiri,Odintsov,Onemli:QE,Carroll:BD} for other
possibilities).  Naturally, the classical Hamiltonian for the
phantom matter becomes unbounded from below. Such unbounded
Hamiltonian essentially leads to a classically unstable system,
as ground state of such system gets pushed to negative infinity.

Apart from the mentioned observational indication of violation of
energy condition, there are in fact theoretical reasons to argue
that some of these energy conditions in general relativity,
should be violated in appropriate regime. One such reason behind
this, is the existence of another important set of theorems, so
called {\em singularity theorems}. These theorems tell us that if
the evolution of a globally hyperbolic spacetime satisfies
Einstein equation and the matter stress-tensor satisfies so
called {\em strong energy condition} then the backward evolution
of such an expanding spacetime is necessarily singular, in a
sense that the spacetime is geodesically incomplete. In the
cosmological context, it implies that if one considers the
backward evolution of an expanding universe with respect to the
{\em coordinate time} (lapse equal to unity) and the matter
content satisfies strong energy condition then the physical
quantities like energy density, spacetime curvature would diverge
within finite time interval. However, the appearance of
singularity in a classical theory is generally considered as an
attempt to extrapolate the classical theory beyond its natural
domain of validity, rather than considering it as a {\em property
of nature}. Near the classical singularity one expects the
evolution of the spacetime to be governed by a quantum theory of
gravity, as classical description signals its own breakdown.
Further, one also believes that a proper theory of quantum
gravity should resolve the singularity that appears in the
classical general relativity.  Naturally, one would naively
expect that the quantum effects of such theory should force the
matter contents to {\em effectively} violate the strong energy
condition when its dynamics is viewed as an evolution of
pseudo-Riemannian spacetime.

In recent years the issues regarding singularities in
cosmological models have been addressed in a rigorous way within
the framework of {\em loop quantum cosmology} (LQC)
\cite{cosmoI,cosmoII,cosmoIII,cosmoIV,IsoCosmo,SemiClass,
LoopCosRev,Bohr,Bojowald:EU}.  The loop quantum cosmology is a
quantization of the cosmological models along the line of a
bigger theory known as {\em loop quantum gravity} (LQG)
\cite{RovelliRev,ThiemannRevI,ThiemannRevII,AshtekarRev}. It has
been shown that the loop quantum cosmology cures the problem of
classical singularities in isotropic model \cite{Sing} as well as
less symmetric homogeneous model \cite{HomoLQC} along with
quantum suppression of classical chaotic behaviour near
singularities in Bianchi-IX models \cite{ChaosSup, BianchiIX}.
Further, it has been shown that non-perturbative modification of
the scalar matter Hamiltonian leads to a generic phase of
inflation \cite{Inflation, GenericInflation} (see
\cite{Singh:ESM} for related discussions on other kinds of
matter). It has been also shown that loop quantum cosmology
induced inflationary scenario can produce {\em scale invariant}
primordial power spectrum as well as observed {\em small
amplitude} for it, without {\em fine tuning} \cite{Hossain:PPS}.
The primordial power spectrum contains a characteristic signature
which is potentially {\em falsifiable} by observations. Further,
it has been argued that the loop quantum cosmology induced
inflationary phase can lead to a secondary standard inflationary
phase \cite{Bojowald:BPI,
Tsujikawa,Bojowald:QMC,Lidsey:ICI,Mulryne:OU}. This follows from
the feature that the in-built inflationary period of loop quantum
cosmology tends to produce favourable initial conditions for an
additional standard inflationary phase. In \cite{Tsujikawa}, the
authors have also studied the possible effects of the above
mechanism on cosmic microwave background (CMB) angular power
spectrum, generated during a standard inflationary phase that
follows the loop quantum cosmology induced inflationary phase and
shown that it can lead to suppression of power in the low CMB
multipoles.  These features crucially depends on a fact that in
loop quantum cosmology the inverse triad (scale factor) operator
\cite{InvScale} whose quantization relies on techniques used in
full theory \cite{Thiemann:QSDV}, has a bounded spectrum. This is
{\em unlike} the classical situation where inverse scale factor
blows up as scale factor goes to zero. However, not being a basic
operator quantization of the inverse scale factor operator
involves quantization ambiguities \cite{Ambig,ICGCAmbig}.

From a quantum mechanical system, generally one obtains
physically relevant quantities by computing physical expectation
values of appropriate physical observables in the relevant
physical states. In loop quantum cosmology, development of the
machinery required to deal with them are still in nascent stage
\cite{cosmoIV,HossainHO,Bojowald:Time,Noui}. Nevertheless, one
can construct an {\em effective} but {\em classical} description
of loop quantum cosmology using WKB techniques. The dynamics of
the effective description is governed by an effective Hamiltonian
\cite{EffectiveHam} along with discreteness corrections
\cite{Banerjee,Vandersloot:HC}. The {\em effective} loop quantum
cosmology incorporates crucial non-perturbative modifications and
has been shown to be generically non-singular as well
\cite{Singh:Bounce,Vereshchagin,GenBounce,Mulryne:EU}. In fact
several important features of loop quantum cosmology, that have
been shown in literature, crucially rely on the effective
classical description. Naturally, in the effective loop quantum
cosmology, the violation of dominant energy condition raises
serious concern. In particular, whether such effective classical
description respects causality. In the cosmological context, any
communication across spatial distances introduces inhomogeneity.
So it is a natural concern to check whether the propagation of
inhomogeneous modes respects causality. Also, whether such
dominant energy condition violating effective description can
ensure stability of the vacuum, as the Hawking-Ellis conservation
theorem no longer guarantees for the same (see also
\cite{Coule:CQC,Coule:QC} for related discussions).

In section II, we briefly review the definitions of relevant
energy conditions used in general relativity. In particular, for
the cosmological context, we discuss the requirements on the
equation of state parameter due to these energy conditions.  In
the next section, we discuss the properties of the equation of
state parameters for a minimally coupled scalar field and also
for the so-called phantom matter model of dark energy. In the
section IV, we study the properties of the effective scalar
matter Hamiltonian. In particular, we show that the kinetic term
due to the modified dynamics, contributes negative pressure even
though it contributes positive energy density. This crucial
feature essentially leads to violation of dominant energy
condition in terms of the equation of state parameter but it also
ensures a bounded (from below) scalar matter Hamiltonian. In the
next section, we derive a modified dispersion relation for the
inhomogeneous modes due to the modified dynamics. Then we show
that the group velocity for the relevant inhomogeneous modes
remains sub-luminal thus ensuring causal propagation across
spatial distances. We also compute the quantum corrections to the
group velocity for a massless free scalar field at large volume.

\section{Energy Conditions in General Relativity}

The energy conditions, often regarded as sacred principles
\cite{Carter}, were mostly postulated to prove several important
theorems in classical general relativity. A few important among
them are the so called {\em singularity theorems} and {\em
conservation theorem}. In this section, we will briefly recall
the definitions of some of these energy conditions. In the
cosmological context, these energy conditions can be essentially
stated in terms of the energy density and its relation to the
pressure component {\em i.e.} the {\em equation of state}
parameter. We will mainly follow the convention of Wald
\cite{Wald}.

\subsection{Weak Energy Condition}

For a given matter stress-energy tensor $T_{\mu\nu}$, the
quantity $T_{\mu\nu}\xi^{\mu}\xi^{\nu}$ physically represents
local energy density for an observer whose 4-velocity is
$\xi^{\mu}$ at a spacetime point. The {\em weak energy condition}
is physically interpreted as the requirement of {\em
non-negativity} for the classical energy density. Naturally, the
weak energy condition is stated as 
\begin{equation}
T_{\mu\nu}\xi^{\mu}\xi^{\nu} \ge 0 ~,
\label{WEC}
\end{equation}
for all time-like $\xi^{\mu}$.  Assuming that the stress-energy
tensor can be diagonalized {\em i.e.} it can be written as
$T_{\mu\nu} := \rho~ t_{\mu} t_{\nu} + P_1~ x_{\mu} x_{\nu} + P_2
~y_{\mu} y_{\nu} + P_3~ z_{\mu} z_{\nu}$ where $\{t^{\mu},
x^{\mu}, y^{\mu}, z^{\mu} \}$ is an orthogonal set of basis and
$t^{\mu}$ is time-like, the weak energy condition requires $\rho
\ge 0$ and $\rho + P_i \ge 0$ for $i=1,2,3$ where $P_i$ is the
{\em principal} pressure. For the homogeneous and isotropic
spacetime these requirements can be conveniently stated in terms
of the equation of state parameter $\omega := P/\rho$ as
$\omega \ge -1$ and the energy density $\rho \ge 0$.

\subsection{Strong Energy Condition}

A crucial requirement on the matter stress-energy tensor, for the
singularity theorems to hold, is that it should satisfy so called
{\em strong energy condition}. This energy condition requires
matter stress-energy tensor to satisfy 
\begin{equation}
T_{\mu\nu}\xi^{\mu}\xi^{\nu} \ge -\frac{1}{2} T  ~,
\label{SEC}
\end{equation}
for all {\em unit} time-like $\xi^{\mu}$.  Assuming diagonal form
of the stress-energy tensor, the strong energy condition requires
$\rho + \sum_{j=1}^3 P_j \ge 0$ and $\rho + P_i \ge 0$ for
$i=1,2,3$. For the homogeneous and isotropic spacetime, these
requirements in terms of the energy density and equation of state 
parameter can be stated as $\rho \ge 0$, $\omega \ge
-\frac{1}{3}$. One may note here that the violation of strong
energy condition which is {\em necessary} for non-singular cosmological
evolution, implies an accelerating phase in its evolution
via Raychaudhuri equation.

\subsection{Dominant Energy Condition}

The Hawking-Ellis conservation theorem requires matter
stress-energy tensor to satisfy so called {\em dominant energy
condition}. This condition requires the local energy density to
be {\em non-negative} for all time-like observer and the local
energy-momentum 4-current {\em i.e} $-T_{\mu\nu}\xi^{\mu}$ to be
future directed, {\em non-spacelike} for all future directed,
time-like $\xi^{\mu}$. So the dominant energy condition is stated
as 
\begin{equation}
T_{\mu\nu}\xi^{\mu}\xi^{\nu} \ge 0 ~~~; ~~~
T_{\mu\nu}\xi^{\nu} T^{\mu}_{\rho}\xi^{\rho} \le 0 ~.
\label{DEC}
\end{equation}
The second requirement can be physically interpreted as the
requirement on matter stress-energy tensor such that the speed of
energy-flow does not exceed the speed of light. Assuming diagonal
form of the stress-energy tensor, the dominant energy condition
requires $\rho \ge |P_i|$ for $i=1,2,3$. In other words, the
energy density is required to {\em dominate} the pressure
components. For the homogeneous and isotropic spacetime, these
requirements can be stated in terms of the equation of state
parameter as $|\omega| \le 1$ and energy density $\rho \ge 0$.

Apart from the above energy conditions, there are few more energy
conditions that can be seen in the literature. For example, so
called {\em null energy condition} requires matter stress-energy
tensor to satisfy $T_{\mu\nu}n^{\mu}n^{\nu} \ge 0 $, for all null
vector $n^{\mu}$.

\section{Classical Scalar Matter Hamiltonian}

In the cosmological scale, our universe appears to be spatially
flat, homogeneous and isotropic with a very good precision. The
invariant distance element in such spacetime (using {\em natural
units} {\em i.e.} $c=\hbar=1$) is given by
Friedmann-Robertson-Walker metric
\begin{equation}
ds^2 ~=~ -dt^2 ~+~ a^2(t)~d{\bf x}^2 ~,
\label{FRWMetric}
\end{equation}
where $a(t)$ is the {\em scale factor}. Clearly, the metric
components do not have any spatial dependence.  In this paper we
will consider a {\em minimally} coupled scalar field as the
matter source. The dynamics of such scalar field is governed by
the action 
\begin{equation} 
S_{\phi} ~:=~ \int d^4 x \sqrt{-g} \mathcal{L} 
~=~ \int d^4 x \sqrt{-g}~ \left[ -\frac{1}{2} g^{\mu\nu}
\partial_{\mu}\phi \partial_{\nu}\phi - V(\phi) \right]
 ~.
\label{ScalarAction}
\end{equation} 
Let us recall that we are mainly interested in studying the
effects on the scalar field dynamics, due to the non-perturbative
modification coming from loop quantum cosmology.  In the
canonical quantization, as in loop quantum cosmology, one treats
Hamiltonian as a basic object that governs the dynamics of the
system. Thus, for our purpose it is necessary to have the
expression for the scalar matter Hamiltonian 
\begin{equation} 
H_{\phi} = a^{-3}~\int d^3 x \left[ \frac{1}{2} {\pi}_{\phi}^2
\right] 
~+~ a ~\int d^3 x \left[\frac{1}{2} {(\nabla \phi)}^2 \right]
~+~ a^3~\int d^3 x \left[ V(\phi) \right] ~,  
\label{SFIHamiltonian}
\end{equation} 
where field {\em momentum density} ${\pi}_{\phi} = a^3 {\dot
\phi}$. In deriving expression (\ref{SFIHamiltonian}), it is
assumed that the background {\em geometry} is homogeneous,
isotropic and described by the metric (\ref{FRWMetric}). However,
we have assumed that the scalar field itself need {\em not} be
homogeneous. This {\em approximation} greatly simplifies the
analysis. Nevertheless, one should keep it in mind that it is
trustworthy as long as the deviation from homogeneity and
isotropy remains small.

In loop quantum cosmology, the geometrical quantities like the
scale factor $a$ here, are represented through corresponding
quantum operators. While deriving effective classical Hamiltonian
from loop quantum cosmology, these geometrical quantities
effectively get replaced by the eigenvalues of their
corresponding quantum operators. The kinetic term of the scalar
matter Hamiltonian (\ref{SFIHamiltonian}) involves inverse powers
of the scale factor. In loop quantum cosmology, the inverse scale
factor operator has a bounded spectrum. Clearly one can see that
the kinetic term of the effective scalar matter Hamiltonian will
involve non-perturbative modifications.

Given an arbitrary inhomogeneous scalar field in a spatially flat
space, one can decompose it in terms its Fourier modes. In this
case, the dynamics of the $k=0$ mode {\em i.e.} the spatially
homogeneous mode will essentially drive the evolution of the
homogeneous background geometry, as the contribution from
non-zero $k$ modes are assumed to be small. So for the purpose of
determining the background evolution, it is sufficient to
consider only the homogeneous mode. In other words, we will
neglect the contribution from the gradient term while evaluating
the background evolution.  Naturally, the scalar matter
Hamiltonian (\ref{SFIHamiltonian}) reduces to 
\begin{equation}
H_{\phi} = p^{-\frac{3}{2}}~\frac{1}{2} p_{\phi}^2 ~+~
p^{\frac{3}{2}}  V(\phi) ~, 
\label{SFBHamiltonian}
\end{equation} 
where $\int d^3 x \sqrt{-g} := a^3 V_0  := p^{\frac{3}{2}}$ and
$p_{\phi} (= V_0 \pi_{\phi})$ is the field {\em momentum}. It is
important to note here that we have absorbed the {\em fiducial}
coordinate volume $V_0$ (of a given finite cell) in the
definition of the variable $p$. In loop quantum cosmology, the
variable $p$ is known as redefined densitized triad and it is one
of the basic phase space variables.  

\subsection{Classical Energy Density and Pressure}

In the Lagrangian formulation, one can obtain the expression for
the general stress-energy tensor by considering the variation of
the action with respect to the spacetime metric. Naturally, one
can use the general expression of the stress-energy tensor, to
obtain the reduced expression for the energy density and the
pressure component for the homogeneous and isotropic spacetime.
On the other hand, in the Hamiltonian formulation such direct
method is not available. However, one can define the expression
for the energy density and the pressure component in terms of the
classical Hamiltonian as
\begin{equation} 
\rho := \frac{1}{2}{\dot\phi}^2 + V(\phi) = 
p^{-\frac{3}{2}}H_{\phi} ~~;~~	
P := \frac{1}{2} {\dot \phi}^2 - V(\phi) = - p^{-\frac{3}{2}}
\left(\frac{2p}{3}\frac{\partial H_{\phi}}{\partial p}\right)~.
\label{ClassicalEP}
\end{equation} 
It may be noted here that the definitions of the energy density
and the pressure (\ref{ClassicalEP}) in terms of the scalar
matter Hamiltonian immediately ensure the matter conservation
equation ${\dot \rho} = -3 \left(\frac{\dot a}{a}\right) (\rho +
P)$ along the classical trajectories.

\subsection{Classical Equation of State}

In the cosmological context, the equation of state parameter is
defined as the ratio of the pressure component to the energy
density as 
\begin{equation} 
\omega ~:=~ \frac{P}{\rho} ~=~ 
\frac{\frac{1}{2}{\dot\phi}^2 - V(\phi)}
{\frac{1}{2}{\dot\phi}^2 + V(\phi)}
\label{ClassicalEOS}
\end{equation} 
For a minimally coupled scalar field, the values of the equation
of state parameter (\ref{ClassicalEOS}) are restricted to be
$|\omega|\le 1$, as the scalar field $\phi$ takes values in the
real line and the potential is required to satisfy $V(\phi)\ge
0$. In other words, the dynamics of a minimally coupled scalar
field always respects the dominant energy condition. Thus, the
Hawking-Ellis conservation theorem vouches for the stability of
the ground state. The stability of the ground state can also be
understood from the property of the scalar matter Hamiltonian
(\ref{SFBHamiltonian}). It is easy to see that the expression of
the scalar matter Hamiltonian (\ref{SFBHamiltonian}) ensures that
it remains bounded from below. This property immediately implies
a classically stable ground state for the system.

\subsection{Phantom Matter Equation of State}

As we have mentioned, in the phantom matter model of dark energy
energy \cite{Caldwell:PM}, one consider a minimally coupled
scalar field but with relatively {\em negative} kinetic term.
Thus, the energy density and the pressure component for the
phantom field are given by 
\begin{equation} 
\rho_{\text{Phantom}} := - \frac{1}{2}{\dot\phi}^2 + V(\phi) ~~;~~	
P_{\text{Phantom}} := - \frac{1}{2} {\dot \phi}^2 - V(\phi)  ~.
\label{PhantomEP}
\end{equation} 
Clearly, the equation of state parameter for the phantom field
$\omega_{\text{Phantom}}( :=  P_{\text{Phantom}} /
\rho_{\text{Phantom}})$  can take value less than $-1$. In other
words, the phantom matter field violates the dominant energy
condition. Naturally, the Hawking-Ellis conservation theorem does
not guarantee for the stability of the ground state. In
particular, using the corresponding Hamiltonian for the phantom
matter, one can easily see that it no longer remains bounded
from below. The unbounded (from below) Hamiltonian immediately
implies that there does not exist a classically stable ground
state for the system.

\section{Effective Scalar Matter Hamiltonian}

In isotropic loop quantum cosmology, the basic phase space
variables are Ashtekar connection and densitized triad.  The
geometrical property of the space is encoded in the densitized
triad $p$ whereas the time variation of geometry is encoded in
the connection. In loop quantum cosmology one redefines
densitized triad to absorb the fiducial coordinate volume
component. This makes the proper volume of the universe
(\ref{FRWMetric}) to be $\int d^3 x \sqrt{-g} ~=~ a^3 V_0 ~=~
p^{\frac{3}{2}}$ \cite{Bohr}. The effective scalar matter
Hamiltonian for the classical system whose dynamics is governed
by the Hamiltonian (\ref{SFBHamiltonian}), is given by
\cite{EffectiveHam}
\begin{equation}
H^{\text{eff}}_{\phi} =
\frac{1}{2} {|\tilde{F}_{j,l}(p)|}^{\frac{3}{2}} {p_{\phi}}^2
~+~ p^{\frac{3}{2}} V(\phi) ~,
\label{EffMatterHam}
\end{equation}
where $\tilde{F}_{j,l}(p)$ is the eigenvalue of the inverse
densitized triad operator $\hat{p^{-1}}$ and is given by
$\tilde{F}_{j,l}(p) = (p_j)^{-1} F_l( p/p_j )$ where $p_j
=\frac{1}{3}\gamma\mu_0 j l_p^2$. The $\mu_0$ is an order of
unity parameter that appears while quantizing the Hamiltonian
constraint operator in loop quantum cosmology \cite{Bohr}. The
$j$ and $l$ are two quantization ambiguity parameters
\cite{Ambig,ICGCAmbig}. The half integer $j$ is related with the
dimension of representation while writing holonomy as
multiplicative operator. The real valued $l$ ($0<l<1$)
corresponds to different, classically equivalent ways of writing
the inverse power of the densitized triad in terms of Poisson
bracket of the basic variables. The function $F_l(q)$ is
approximated as \cite{BianchiIX}
\begin{eqnarray}
F_l(q)&=& \left[ \frac{3}{2(l+2)(l+1)l} \left( ~
(l+1) \left\{ (q + 1)^{l+2} - \left. |q - 1|^{l+2} \right\} ~-~ 
\right. \right. \right. \nonumber \\
& & ~~~(l+2) q \left\{ (q + 1)^{l+1} - \right. 
 \left. \left. \left. \sgn(q - 1) |q - 1|^{l+1} \right\}
~ \right) ~\right]^{\frac{1}{1-l}} \nonumber \\
& \rightarrow &  q^{-1}  ~~~~~~~~~~~~~~~~~(q \gg 1) \nonumber \\
& \rightarrow &  \left[ \frac{3 q}{l+1}\right]^{\frac{1}{1-l}}
~~~~~~~(0 < q \ll 1) ~. 
\label{InvSF}
\end{eqnarray}
It is clear from the expression (\ref{InvSF}) that for the large
values of the densitized triad {\em i.e.} in large volume one
recovers the expected classical behaviour for the inverse
densitized triad. The quantum behaviour is manifested for smaller
values of the densitized triad. Here the meaning of large or
small values of the triad $p$ is determined necessarily by the
values of $p_j$. We will follow this convention throughout the
paper unless explicitly stated.

\subsection{Effective Energy Density and Pressure}

In this paper, we are interested in studying the effects on the
{\em energy conditions} due to the non-perturbative modification
coming from loop quantum cosmology and its further implications.
In the cosmological context, the energy conditions are stated in
terms of the energy density and its relation to the pressure {\em
i.e.} the equation of state parameter. In loop quantum cosmology,
one obtains non-perturbative modification at the level of the
effective Hamiltonian but {\em not} at the level of an effective
action. This prevents one to directly obtain the expression of
the {\em effective} stress-energy tensor. On the other hand, in
classical general relativity the energy conditions are defined in
terms of the stress-energy tensor. Naturally, the issue of energy
conditions violation in the effective dynamics, is crucially
related to the definitions of the effective energy density and
pressure. In the classical situation we have seen that it is
possible to write down the reduced standard expressions of the
energy density and the pressure (\ref{ClassicalEP}) purely in
terms of the reduced Hamiltonian.  These definitions of the
energy density and the pressure immediately ensure the matter
conservation equation along the classical trajectories.
Naturally, one can use the same definitions for the effective
energy density and the pressure just replacing the standard
Hamiltonian in terms of the effective Hamiltonian. So we define
the {\em effective} energy density and the {\em effective}
pressure, following the definitions of classical energy density
and pressure (\ref{ClassicalEP}), as
\begin{equation} \rho^{\text {eff}} :=
p^{-\frac{3}{2}}H^{\text{eff}}_{\phi} ~~;~~	P^{\text{eff}} :=
- p^{-\frac{3}{2}} \left(\frac{2p}{3}\frac{\partial
H^{\text{eff}}_{\phi}}{\partial p}\right)~.  
\label{EffectiveEP}
\end{equation}
It is worth pointing out that to define the effective energy
density and the pressure, one could have proceeded as done in
\cite{EffectiveHam}.  In this approach one first obtains the
Hamilton's equations of motion for the matter degrees of freedom
as well as the  gravitational degrees of freedom. Then one
rewrites these equations of motion, by suitable manipulations
such that a part of these equations matches with the
gravitational part of the standard Friedman equation and the
Raychaudhuri equation. In the next step, one then reads off the
expressions for effective energy density and the pressure by
comparing with standard equations. These expressions of the
energy density and the pressure agree with the definitions
(\ref{EffectiveEP}) when the contributions due to the
non-perturbative modification of the {\em gravity sector} become
negligible. Since the effective Hamiltonian description is
strictly valid in the region where background geometry is
essentially classical {\em i.e.} non-perturbative modification of
geometry is negligible. Clearly, in such situation these two set
of definitions agree with each other. It is important to
emphasize here that although the non-perturbative modification of
the {\em gravity sector} becomes negligible in the region of
interest but the non-perturbative modification of the {\em matter
sector} can still survive. In fact we are interested in studying
the effects of non-perturbative modification of the scalar matter
dynamics. 

\subsection{Effective Equation of State}

Having known the expressions of the effective energy density and
the pressure (\ref{EffectiveEP}), one can easily define the
effective equation of state parameter $\omega^{\text{eff}} :=
P^{\text {eff}}/\rho^{\text {eff}}$. The evolution of the
effective equation of state parameter depends on the effective
Hamiltonian. However, as shown in \cite{GenericInflation}, one
can eliminate the explicit appearance of the effective
Hamiltonian and can express the effective equation of state
parameter in terms of the classical equation of state parameter
$\omega$, as
\begin{equation}
\omega^{\text{eff}} =  -1 + \frac{(1+\omega)p^{\frac{3}{2}}
[\tilde{F}_{j,l}(p)]^{\Case{3}{2}} \left(1 -
\frac{p}{\tilde{F}_{j,l}(p)} \frac{d \tilde{F}_{j,l}(p)}{d p}
\right)}
{(1+\omega)p^{\frac{3}{2}}[\tilde{F}_{j,l}(p)]^{\frac{3}{2}}  +
(1-\omega)} ~.	
\label{EffectiveEOS}
\end{equation}
Using the expression (\ref{InvSF}), it is easy to see that for
the large values of the densitized triad $p$, where one expects
the quantum effects to be small, $\omega^{\text{eff}} \simeq
\omega$.  On the other hand, for small values of $p$,
$\omega^{\text{eff}}$ differs from the classical $\omega$
dramatically.  Using the small volume (small triad) expression of
the inverse densitized triad (\ref{InvSF}), one may note that the
effective equation of state satisfies $(\omega^{\text{eff}} + 1)
< 0$, for all allowed values of the ambiguity parameter $l$.
Let's recall that in terms of equation of state parameter, the
{\em weak} energy condition requires $(\omega+1)\ge 0$, the {\em
strong} energy condition requires $(\omega+\frac{1}{3})\ge 0$ and
the  {\em dominant} energy conditions requires $|\omega| \le 1$.
So it is clear that in loop quantum cosmology, the effective
equation of state parameter violates all of these energy
conditions due to the non-perturbative modifications.

\subsection{Kinetic Contribution to Pressure}

The allowed values for the classical equation of state
(\ref{ClassicalEOS}), are restricted to be $|\omega|\le 1$.
Naturally, it is an important question to ask how is it then
possible for the effective equation of state to take values less
than $-1$, instead of the facts that in both cases one begins
with a {\em minimally} coupled scalar field and uses the same
definition for the equation of state parameter in terms of their
corresponding Hamiltonian. The answer to this question lies in
the fact that in effective loop quantum cosmology, although one
begins with a {\em standard minimally} coupled scalar field but
for small volume this coupling gets altered dramatically. The
effective coupling remains {\em minimal} in a sense that it
couples only through the geometrical variables but {\em not}
through curvatures. However, it is clear that the gravity
coupling to the scalar matter no longer remains {\em standard}
minimal coupling as the spectrum of the inverse triad
operator differs from the classical expression dramatically for
small volume. To understand this issue better, let us have a look
at the contributions due to the kinetic term to the pressure
component $P_{\text{KE}}$ for both cases 
\begin{equation} P_{\text{KE}} = -~ p^{-\frac{3}{2}}
\left[\frac{p}{3} p_{\phi}^2\right] \frac{\partial} {\partial p}
\left[ p^{-\frac{3}{2}}\right] ~ ~;~ ~
P^{\text{eff}}_{\text{KE}} = -~ p^{-\frac{3}{2}}
\left[\frac{p}{3} p_{\phi}^2~\right] \frac{\partial} {\partial p}
\left[ {|\tilde{F}_{j,l}(p)|}^{\frac{3}{2}} \right]~.
\label{KEPressure} \end{equation} It is evident from the equation
(\ref{KEPressure}) that in the standard case, the kinetic term
contributes {\em positive} pressure. This is what one would
intuitively expect from our understanding of ordinary
thermo-dynamical system. However, in the effective loop quantum
cosmology, using the expression of the inverse densitized triad
(\ref{InvSF}), it is easy to see that the kinetic term
contributes {\em negative} pressure for small volume even though
for large volume it contributes positive pressure like in
standard case. This crucial `extra' negative pressure from the
kinetic term is what essentially leads the effective equation of
state to violate dominant energy condition. Clearly, the bounded
spectrum of the inverse densitized triad plays a major role in
this.

On the other hand, in the phantom matter model of dark energy, to
obtain the values of the equation of state parameter to be less
than $-1$, one makes the kinetic term relatively {\em negative}
by hand. This step essentially forces the kinetic term to
contribute negative pressure. However, it also leads the kinetic
term to contribute negative energy density. This step essentially
jeopardise energy density expression as its {\em positivity} is
no longer remain guaranteed. Clearly, a relatively {\em negative}
kinetic term in the scalar matter Hamiltonian, makes it unbounded
from below. In other words, the ground state of such system gets
pushed to negative infinity. Naturally, naive quantization of
such system can lead to a catastrophic decay of vacuum
\cite{Carroll:EOS}. On contrary, in the effective loop quantum
cosmology scenario, the kinetic term gives negative contribution
only in the pressure expression but not in the energy density
expression.  Thus, although the equation of state parameter in
effective loop quantum cosmology violates dominant energy
condition but it also necessarily ensures the {\em positivity} of
the energy density. It is also evident from the expression of the
scalar Hamiltonian (\ref{EffMatterHam}) that it remains bounded
from below signifying a stable ground state. 

\subsection{Example: Massive Scalar Field}

Now we take an explicit example to illustrate the dynamics of the
scalar field at small volume regime where non-perturbative
modification plays a significant role. For simplicity, we
consider the dynamics of a massive free scalar field. In other
words, the scalar potential is consist of only the mass term {\em
i.e.} $V(\phi) = \frac{1}{2} m_{\phi}^2 \phi^2$. To simplify
further, we choose the value of the ambiguity parameter to be $l
\rightarrow 0+$. With these assumptions the effective matter
Hamiltonian for small volume becomes
\begin{equation}
H^{\text{eff}}_{\phi} \simeq p^{\frac{3}{2}}~
\left[\frac{1}{2}~\left(3^{\frac{3}{2}}  p_j^{-3}\right)
p_{\phi}^2 ~+~  \frac{1}{2} m_{\phi}^2 \phi^2 \right] ~.
\label{EMHExample}
\end{equation}
Using the Hamilton's equations of motion, one can obtain
analytical solutions for the field equations, given by
\begin{equation}
\phi = {\sqrt \frac{2\bar{\rho}}{m_{\phi}^2}}~ 
\text{sin} \left( \alpha p^{\frac{3}{2}} + c_1 \right) ~;~
p_{\phi}  = {\sqrt \frac{2\bar{\rho}}{( 3^{\frac{3}{2}} p_j^{-3})}}
~ \text{cos} \left( \alpha p^{\frac{3}{2}} + c_1 \right) ~,
\label{ExampleSol}
\end{equation}
where $\alpha = {\sqrt\frac{ (3^{\frac{3}{2}}
p_j^{-3})(m_{\phi}^2)}{24\pi G \bar{\rho}}}$, $\bar{\rho}$ and
$c_1$ are two constants of integration. Using the field solutions
(\ref{ExampleSol}), one can easily see that along any trajectory
$H^{\text{eff}}_{\phi} \simeq p^{\frac{3}{2}}~\bar{\rho}$. One
may note here that the energy density contribution due to the
scalar field dynamics effectively looks a like contribution from
a cosmological constant. The constant of integration $\bar{\rho}$
physically corresponds to the energy density during its
evolution. This also implies an exponential inflationary phase.
This is of course expected behaviour, as the effective equation
of state parameter in loop quantum cosmology generically becomes
$\omega^{\text{eff}}\approx -1$ at small volume
\cite{GenericInflation}. This simple example clearly shows that
classical dynamics of the system is essentially stable, as we
have argued for a general system with the modified scalar field
dynamics.

\section{Propagation of Inhomogeneous Modes}

We have mentioned earlier that the second part of the dominant
energy condition requires the speed of energy propagation not to
exceed the speed of light. Naturally, the violation of dominant
energy condition also raises the concern, whether such system can
prohibit super-luminal flow of energy. In other words, whether
such system can respect causality. In classical cosmology, one
begins by postulating so called {\em cosmological principle i.e.}
on large scale {\em there is neither a preferred direction nor a
preferred place} in our universe. This principle is imposed by
assuming that on cosmological scale our universe is spatially
homogeneous and isotropic. The strict imposition of spatial
homogeneity will prohibit any kind of spatial flow of energy as
it will violate spatial homogeneity. However, this assumption
undoubtedly is an idealisation and is made to rather simplify
background dynamics.  Naturally, if we want to allow some kind of
spatial flow of energy then we must relax the spatial
homogeneity. While relaxing this assumption nevertheless one
should be careful so that we can still use the available
machinery of the cosmological set-up.  This is generally achieved
by considering the deviation from spatial homogeneity to be
small. It is worth pointing out that small spatial inhomogeneity
in the matter field configuration will also lead to small
inhomogeneity in the background geometry.  For simplicity,
however, we will treat the background geometry as homogeneous.

\subsection{Modified Klein-Gordon Equation}

We have seen earlier that the kinetic term of the scalar matter
Hamiltonian gets non-perturbative modification, as its classical
expression involves inverse powers of densitized triad. The
effective scalar matter Hamiltonian, obtained as outlined in
\cite{Hossain:PPS}, is given by
\begin{equation} 
H^{\text{eff}}_{\phi} = V_0 {|\tilde{F}_{j,l}(p)|}^{\frac{3}{2}} 
\int d^3 x \left[ \frac{1}{2} {\pi}_{\phi}^2 \right] 
~+~ V_0^{-\frac{1}{3}} p^{\frac{1}{2}} \int d^3 x 
\left[\frac{1}{2} {(\nabla \phi)}^2 \right]
~+~ V_0^{-1} p^{3/2} \int d^3 x \left[ V(\phi) \right] ~.  
\label{EffectiveHamiltonian}
\end{equation} 
One may note here that we have now kept the gradient term in the
effective Hamiltonian. The gradient term was neglected earlier
while computing background evolution, as one assumes that the
background evolution is mainly determined by the homogeneous and
isotropic contribution of the matter Hamiltonian. It is worth
pointing out here that the gradient term of the equation
(\ref{EffectiveHamiltonian}) having the correct sign, the
corresponding dynamics does not suffer from the so called
gradient instability \cite{Hsu:GI,Buniy}, another pathological
feature of the phantom matter models.
Using the 
Hamilton's equations of motion for the effective Hamiltonian
(\ref{EffectiveHamiltonian}), one can derive the corresponding
{\em modified} Klein-Gordon equation
\begin{equation} 
\ddot \phi ~-~ 3 
\left(\frac{p~\tilde{F}_{j,l}'(p)}{ \tilde{F}_{j,l}(p)} \right)
\left(\frac{\dot a}{a}\right)
\dot \phi ~+~ 
{|\tilde{F}_{j,l}(p)|}^{\frac{3}{2}} p^{\frac{3}{2}} 
 \left( - \frac{\nabla^2 \phi}{a^2} + V'(\phi) \right) ~=~ 0 ~,
\label{MKGEquation}
\end{equation} 
where $\tilde{F}_{j,l}'(p) \equiv \frac{d \tilde{F}_{j,l}(p)}{d
p}$. Using the expression for the spectrum of the inverse triad
(\ref{InvSF}), it is easy to see that the modified Klein-Gordon
equation (\ref{MKGEquation}) reduces to the standard Klein-Gordon
equation at large volume.

In a given spatially flat spacetime background, an inhomogeneous
scalar field can be decomposed in terms of its Fourier modes. The
dynamics of the $k=0$ mode {\em i.e.} the spatially homogeneous
mode essentially drives the evolution of the background geometry
as the contributions from non-zero $k$ modes are assumed to be
small. However, as we have argued that to study the energy
propagation across spatial distance in the cosmological
background, it is essential to consider the dynamics of
inhomogeneous modes {\em i.e.} non-zero $k$ modes. The Fourier
decomposition of the inhomogeneous scalar field is defined as
\begin{equation} 
\phi({\bf x}, t) ~=~ \int \frac{d^3 {\bf k}}{{(2\pi)}^3}
\left[ \phi_k(t)~ e^{i {\bf k \cdot x}} \right] ~, 
\label{FourierModes}
\end{equation}
where $\phi_k(t)$ are the Fourier components. For simplicity, we
will consider the dynamics of a massless free scalar {\em i.e.}
we will assume $V(\phi)=0$. Using the modified Klein-Gordon
equation (\ref{MKGEquation}) and the equation
(\ref{FourierModes}), one can derive the modified equation for
the Fourier modes 
\begin{equation} 
\ddot \phi_k(t) ~-~ 3 
\left(\frac{p~\tilde{F}_{j,l}'(p)}{ \tilde{F}_{j,l}(p)} \right)
\left(\frac{\dot a}{a}\right) 
\dot \phi_k(t) ~+~
{|\tilde{F}_{j,l}(p)|}^{\frac{3}{2}} p^{\frac{3}{2}} 
~\left(\frac{k^2}{a^2}\right)
\phi_k(t) ~=~ 0 ~.
\label{ModifiedFMEquation}
\end{equation} 
In the small volume regime where the spectrum of the inverse
triad operator can be approximated as $\tilde{F}_{j,l}(p) \sim
p^{\frac{1}{1-l}}$ and the effective equation of state parameter
as $\omega^{\text{eff}} \approx -1$, one can obtain an analytical
solution for the equation (\ref{ModifiedFMEquation})
\cite{Hossain:PPS}, given by 
\begin{equation} 
\phi_k(t) ~=~ \eta^{(1+\frac{1}{2n})}
\left[ A_{(k,n)} J_{-(1+\frac{1}{2n})}(k\eta)
+  B_{(k,n)} J_{(1+\frac{1}{2n})}(k\eta) \right]  ~,
\label{FMSolution}
\end{equation} 
where $J_n(x)$ are the Bessel functions, $A_{(k,n)}$ and
$B_{(k,n)}$ are two constants of integration corresponding to
{\em second} order differential equation. The variable $\eta$ is
defined as $d \eta := a^{-n} d t$, where the parameter $n =
-\frac{1}{2}(1+ \frac{3}{1-l})$. In loop quantum cosmology
allowed values for the ambiguity parameter $l$ is $(0<l<1)$.
Naturally, the new parameter $n$ takes values as
($-\infty<n<-2$). The argument of the Bessel function $k\eta$
can be conveniently expressed in terms the scale factor as $k\eta
= \frac{k}{H a} \left(\frac{a^{1-n}}{-n}\right)$, where $H(\equiv
\frac{\dot a}{a})$ is the Hubble parameter.

At first let's study the {\em large} wavelength ($k \rightarrow
0$) behaviour of the general solution (\ref{FMSolution}). For the
general solution (\ref{FMSolution}) when both constants of
integration $A_{(k,n)}$ and $B_{(k,n)}$ are present then using
the asymptotic form of the Bessel function $J_{m}(x) \approx
\frac{1}{\Gamma(1+m)}(\frac{x}{2})^m$ for $x<<1$, it is easy to
see that $\phi_k(t)$ becomes approximately constant and becomes
proportional to $A_{(k,n)}$. For the special case when
$A_{(k,n)}$ is identically zero then $\phi_k(t)$ remains time
dependent but its time dependence is {\em non-oscillatory}. These
features of the Fourier modes $\phi_k(t)$ can also be seen
directly from the differential equation
(\ref{ModifiedFMEquation}). For the larger wavelength modes the
third term in the equation (\ref{ModifiedFMEquation}) can be
neglected. The approximated second order differential equation
then admits a {\em constant} solution and a {\em non-oscillatory}
time-dependent solution, as expected. Clearly, the second term
which is a (anti)friction term, plays a major role for the larger
wavelength modes. Since, our main interest is to study the energy
transmission across spatial distances then clearly the larger
wavelength modes are not relevant for this purpose. On the other
hand, for smaller wavelength ($k \rightarrow \infty$) modes, the
general solution become {\em oscillatory}, as the asymptotic form
of the Bessel function is $J_m(x) \approx {\sqrt \frac{2}{\pi x}}
\text{cos}\left(x - \frac{m\pi}{2} - \frac{\pi}{4}\right)$ for
$x>>1$. Naturally, the smaller wavelength modes are the potential
{\em carriers} for the energy transmission across spatial
distances. For smaller wavelength modes, the effects of the
(anti)friction term is negligible. Thus, for simplicity we will
neglect the (anti)friction term in the equation
(\ref{ModifiedFMEquation}) for further analysis. The information
regarding assumed small inhomogeneity are encoded in the
amplitudes of the mode functions $\phi_k(t)$. Since propagation
speed of linear waves  does not depend on their amplitudes, the
causal properties of the propagating inhomogeneous modes are
quite insensitive to the exact details of their amplitudes.

\subsection{Modified Dispersion Relation}

In the cosmological context, any spatial transmission of energy
will introduce inhomogeneity. So to investigate causality of the
system, it is natural to study the {\em group velocity} for the
inhomogeneous modes. One may recall that in a medium where
absorption (friction) or emission (anti-friction) is small, the
{\em group velocity} essentially determine the speed of signal
propagation \cite{Brillouin}. To compute the group velocity it is
convenient to find out the relation between its frequency and
wave-number {\em i.e.} the {\em dispersion relation}. Using the
governing equation for the inhomogeneous modes
(\ref{ModifiedFMEquation}), neglecting the (anti)friction term,
and making the ansatz $\phi_k(t) \sim e^{i \tilde{\omega} t}$, one can
easily derive the {\em modified} dispersion relation in effective
loop quantum cosmology as
\begin{equation} 
{\tilde{\omega}}^2 ~\approx~ 
{|\tilde{F}_{j,l}(p)|}^{\frac{3}{2}} p^{\frac{3}{2}} 
~\left(\frac{k^2}{a^2} \right) ~.
\label{DispersionRelation}
\end{equation} 
In the classical situation `inverse triad' is just the inverse of
triad {\em i.e} $p\times\tilde{F}_{j,l}(p) = 1$.  The dispersion
relation (\ref{DispersionRelation}) then becomes same as the
standard Minkowskian dispersion relation between frequency and
{\em physical} wave number $(k/a)$. In loop quantum cosmology the
spectrum of the inverse triad operator is bounded. Hence the
dispersion relation in effective loop quantum cosmology differs
dramatically for small volume compared to the standard dispersion
relation. 

It is worth emphasising that the modification in the dispersion
relation that is being studied here, arises because of the
bounded spectrum of the inverse triad. This modification is
distinct from the different types of modification generally
considered in the literature. For example, in the context of
quantum gravity scenario
\cite{Bojowald:LI,Gambini:QS,Alfaro:QGCNP,Alfaro:LQGLP,
Sahlmann:QFTCS} or in the context of trans-Planckian inflationary
scenario \cite{Brandenberger:TPI,Niemeyer}, one considers
modification of standard dispersion relation by introducing
appropriate {\em non-linearity}.

\subsection{Group velocity}

It is worthwhile to emphasize here that the group velocity
determines the speed of signal propagation only if the absorption
or amplification of the signal remains small. In other words,
`signal transmission' makes sense only if the original signal
reaches its target without major {\em distortion} while
propagation (see \cite{Brillouin} for related discussion).  In
the effective loop quantum cosmology scenario, we have argued
that the relevant modes for energy transmission across spatial
distances, are the smaller wavelength modes and for these modes
the (anti)friction term plays very little role in their
evolution. Using the dispersion relation
(\ref{DispersionRelation}), one can easily compute the {\em group
velocity} for the inhomogeneous modes as
\begin{equation} 
v_g := \frac{d \tilde{\omega}}{d(k/a)} ~=~ 
{|\tilde{F}_{j,l}(p)|}^{\frac{3}{4}} p^{\frac{3}{4}} ~.
\label{GroupVelocity}
\end{equation} 
In the classical situation {\em right hand side} of the
expression (\ref{GroupVelocity}) is identically equal to unity.
Physically, this implies that for the massless free scalar field,
the inhomogeneous modes transmit signals at the speed of light.
However, in the effective loop quantum cosmology it is no longer
the case. Using the expression for the spectrum of the inverse
triad (\ref{InvSF}), it is easy to see that in the small
volume regime, the speed of signal propagation is in fact much
{\em slower} than the speed of light (in classical vacuum). The
group velocity for the inhomogeneous modes gradually increases
and approaches the speed of light towards the end of the
non-perturbatively modified dynamics.

It is worth emphasising here that the actual spectrum of the
inverse scale factor operator is fundamentally
non-differentiable. However, to study the qualitative
consequences of it within an effective analysis, one uses a
peace-wise analytic function $F_l(q)$ (\ref{InvSF}) which
approximates the spectrum of the inverse scale factor operator.
This is a good approximation provided the scale $p_j$ is
sufficiently large. However, being peace-wise analytic this
approximation is good as long as ($p << p_j$) or ($p >> p_j$) but
not near the transition regime, as the approximation function
$F_l(q)$ (\ref{InvSF}) is not analytic at $q=1$ ($q=p/p_j$).  So
the governing equation of the mode functions
(\ref{ModifiedFMEquation}) which involves $F_l(q)$ as well as its
derivative, is not defined near $p=p_j$. Thus, the derivation and
the subsequent expression of the group velocity
(\ref{GroupVelocity}) are valid as long as ($p << p_j$) or ($p >>
p_j$) but not in the neighbouring regime of $p=p_j$. However,
there still exist a significant small volume regime even
excluding the regime near $p=p_j$, as the validity of
approximation for the spectrum of the inverse scale factor
operator, requires $p_j$ to be large.

Thus, in effective loop quantum cosmology although
non-perturbatively modified dynamics violates dominant energy
condition in terms of the equation of state parameter but the
underlying modified dynamics restricts the group velocity for the
inhomogeneous modes to remain {\em sub-luminal}. In the
cosmological context we have argued that any spatial transmission
of energy will introduce spatial inhomogeneity. Here in the
effective loop quantum cosmology, we have shown that the group
velocity for the inhomogeneous modes remains sub-luminal due to
the non-perturbative modification. Clearly, in effective loop
quantum cosmology, non-perturbatively modified dynamics of a
minimally coupled scalar field respects {\em causality}. The
violation of dominant energy condition is essentially dictated by
the $k=0$ mode but this mode is not relevant for the purpose of
signal transmission across spatial distances.

It is worth pointing out that the `speed of light' here is meant
to imply the speed of electromagnetic wave propagation in the
classical vacuum that determines the causal structure of the
spacetime. This is important to emphasize because in loop quantum
cosmology, one expects to get similar non-perturbative
modification even to the electromagnetic wave propagation. Then
the actual speed of light in the effective loop quantum cosmology
itself may become slower compared to the speed of light in the
classical vacuum. Intuitively, one may consider the small volume
{\em effective} background geometry, coming from loop quantum
cosmology, as a {\em refractive} medium with a value of the {\em
group index} $n_g (\equiv c/v_g)$ is greater than unity. In this
context, the group index is same as the {\em refractive index},
as the {\em phase velocity} is same as the group velocity.

\section{Quantum Corrections to group velocity at Large Volume}

Using the spectrum of the inverse triad operator (\ref{InvSF}),
it is easy to see that although at large volume the leading term
is just the inverse of triad but there are sub-leading terms also
in its expression. Naturally, in the effective loop quantum
cosmology, the group velocity for the inhomogeneous modes is not
exactly equal to unity even at the large volume. Using the
expression (\ref{InvSF}), for a massless free scalar field, one
can compute the group velocity with quantum corrections as
\begin{equation} 
v_g  ~\simeq~ \left[1 + \frac{3(2-l)}{40}
\left(\frac{p_j^2}{p^2}\right) \right]~.
\label{GroupVelocityQC}
\end{equation} 
It is clear from the expression (\ref{GroupVelocityQC}) that the
corrections to the group velocity at large volume is {\em
extremely} small but positive as ($0<l<1$). The group velocity
becomes equal to unity as the volume of the system goes to
infinity. To have some numerical estimate of this finite volume
quantum correction, let us choose say $p_j \sim 10^5 ~ l_p^2$.
The observed size of universe today is ${\sqrt p}$ $\sim
10^{60}~l_p$. Then the correction to the group velocity due to
modified spectrum of the inverse triad operator, today is $\sim
10^{-231}~$! It is extremely unlikely that such small correction
will have any significant effect. Even for the cosmological
context (time scale $\sim 10^{17}$ sec) such small deviation of
group velocity, may be completely irrelevant.

\section{Discussions}

To summarize, in effective loop quantum cosmology,
non-perturbatively modified dynamics of a minimally coupled
scalar field violates {\em weak}, {\em strong} and {\em dominant}
energy conditions when they are stated in terms of equation of
state parameter. The violation of strong energy condition
although helps to have non-singular evolution by evading the
singularity theorems but the violation of weak and dominant
energy conditions raises concern. In classical general
relativity, these energy conditions are used to prohibit
super-luminal flow of energy and to ensure the stability of
classical vacuum via the Hawking-Ellis conservation theorem.
Naturally, the violation of these energy conditions in terms of
{\em effective} equation of state parameter, raises concern about
the causality and the stability of the system. In this paper, we
have shown that although at face value these energy conditions
are violated, underlying modified dynamics in effective loop
quantum cosmology nevertheless ensures positivity of energy
density, as scalar matter Hamiltonian remains bounded from below.
Considering the modified dynamics for the inhomogeneous modes, we
have shown that group velocity for the relevant modes remains
sub-luminal in small volume regime, thus ensuring causal
propagation across spatial distances. We have also computed the
large volume quantum corrections to the group velocity of the
inhomogeneous modes for the massless free scalar field.

Now, let us try to understand the physical phenomena behind this
rather unusual feature of the non-perturbatively modified
dynamics. In the case of classical dynamics of a minimally
coupled scalar field, the values of the equation of state
parameter are restricted to be $|\omega| \le 1$. However, in the
case of modified dynamics, the effective equation of state can
take values less than $-1$. This is rather surprising given the
facts that one begins with a minimally coupled scalar field and
uses the same definition of equation of state for both the cases.
This `anomalous' behaviour follows from the fact that at the small
volume, non-perturbatively modified gravity becomes {\em
repulsive} although it remains {\em attractive} for the large
volume. This feature can be easily seen by considering a
classical trajectory of a massless free scalar field. The
non-perturbatively modified scalar matter Hamiltonian, along any
trajectory, increases with the increasing scale factor for small
volume but decrease for large volume. Naturally, the
gravitational Hamiltonian, to satisfy the Hamiltonian constraint
($H_{\phi} + H_{grav}=0$), must decrease with increasing scale
factor for small volume. Later, in the large volume it starts
increasing with increasing scale factor. This immediately implies
that modified gravitational interaction is repulsive for small
volume whereas for large volume, as one expects, it is
attractive. This repulsive nature of the gravitational
interaction manifest itself through the non-standard gravity
coupling to the scalar matter Hamiltonian via bounded spectrum of
the inverse triad operator.

For a standard minimally coupled scalar field, the kinetic term
contributes positive pressure. Of course, this is what one would
intuitively expect from our understanding of ordinary
thermo-dynamical system. However, in the effective loop quantum
cosmology, the kinetic term contributes negative pressure for
small volume even though for large volume it contributes positive
pressure like in standard case. This crucial `extra' negative
pressure from the kinetic term is what essentially leads the
effective equation of state to violate dominant energy condition.
Clearly, the bounded spectrum of the inverse densitized triad
plays a major role in this. On the other hand, in phantom matter
model of dark energy, to obtain the values of the equation of
state parameter to be less than $-1$, one makes the kinetic term
relatively negative by hand. This change of sign essentially
forces the kinetic term to contribute negative pressure. However,
it also leads the kinetic term to contribute negative energy
density. This step badly affects the energy density expression,
as its positivity is no longer certain. In other words, a
relatively negative kinetic term in the scalar matter
Hamiltonian, makes it unbounded from below. This implies that the
system does not have a stable classical ground state.  On
contrary, in the effective loop quantum cosmology scenario, the
kinetic term gives negative contribution only in the pressure
expression but not in the energy density expression. Thus,
although the equation of state parameter in effective loop
quantum cosmology violates dominant energy condition but it
necessarily ensures the positivity of the energy density. It is
also evident from the expression of the scalar Hamiltonian
(\ref{EffMatterHam}) that it remains bounded from below,
signifying a stable classical ground state.

The bounded spectrum of the inverse triad (scale factor) operator
plays the central role in violating the energy conditions.  The
violation of energy conditions although leads to a generic
inflationary phase and allows to have a non-singular evolution
but it also makes the causality and the stability of the system
uncertain. However, as shown in this paper, the same bounded
spectrum in fact acts as a {\em saviour} to ensure the causality
and the stability of the system. It is worth pointing out that
the quantization of the inverse triad was {\em not} invented to
obtain the bounded spectrum such that these physical features
follow. Rather it was quantized following the techniques used in
the full theory of loop quantum gravity. The quantization of the
inverse triad involves ambiguities but these crucial features are
insensitive to their precise values. It may be worth emphasising
that although the exercise presented here is not directly related
with the dark energy scenario, one may learn an important lesson
from here that if one wants to construct a dominant energy
condition violating yet well behaved scalar field model of dark
energy then one should look beyond the standard minimal coupling.

It is now important to discuss some subtleties of the analysis
presented here. In classical general relativity, the definitions
of the energy conditions are generally covariant. However, in the
cosmological context, the energy conditions are stated with
respect to a preferred frame namely the so-called {\em comoving
frame}. Thus, one must be careful while interpreting the results
in more general context.  Secondly, in the Lagrangian formulation
one obtains the reduced expression of the energy density and
pressure for the homogeneous and isotropic spacetime, using a
generally covariant expression of the stress-energy tensor. In
the Hamiltonian formulation such a spacetime covariant method is
not available. Naturally, one needs to define the expression of
energy density and pressure, in terms of the scalar matter
Hamiltonian. In the classical situation although they are
equivalent but with the non-trivial quantum corrections this
issue is rather subtle. In the analysis presented here, we have
assumed the background geometry as homogeneous although we have
allowed the scalar field living in it to become inhomogeneous.
This approximation is trustworthy as long as the deviation from
the homogeneity remains sufficiently small. Further, we have
considered the non-perturbative modification of the kinetic term
only. Using slightly different quantization strategy, one could
obtain a factor of `triad times inverse triad' also in the
gradient term. However, such modification would change only the
{\em quantitative} nature of the results shown here but not the
{\em qualitative} nature. Naturally, the features of the
non-perturbatively modified dynamics shown here, are robust under
this quantization ambiguity.

\begin{acknowledgments}
I thank Ghanashyam Date and Martin Bojowald for careful reading,
critical comments and insightful suggestions on the manuscript. I
thank Ghanashyam Date, Parampreet Singh, Romesh Kaul for helpful,
illuminating discussions.
\end{acknowledgments}


\end{document}